\documentclass[11pt,showpacs,english,superscriptaddress]{revtex4}
\usepackage{graphicx}

\newcommand{\beq}[1]{
\marginpar{\small\textsf{#1}}
\begin{equation}\label{#1}}
\newcommand{\bea}[1]{
\marginpar{\small\textsf{#1}}
\begin{eqnarray}\label{#1}}

\begin{document}

\def\barr{\begin{array}}
\def\earr{\end{array}}

\begin{flushright}
\date{July 8, 2005}
\end{flushright}
\vspace{2cm}
\title{Nucleon Form Factors from Generalized Parton Distributions}
\author{M. Guidal}
\affiliation{Institut de Physique Nucl\'{e}aire Orsay, F-91406 Orsay, France} 
\author{M. V. Polyakov}
\affiliation{Petersburg Nuclear Physics Institute, Gatchina,
    St.~Petersburg 188300, Russia}
\affiliation{Universit\'e  de Li\`ege au Sart Tilman,
                   B-4000 Li\`ege 1 Belgium}
\author{A. V. Radyushkin}
\affiliation{Physics Department, 
Old Dominion University, Norfolk, VA 23529, USA} 
\affiliation{Thomas Jefferson National Accelerator Facility, Newport News, VA 23606, USA}  
\affiliation{Joint Institute for Nuclear Research, 141980 Dubna,  Russia} 
\author{M. Vanderhaeghen}
\affiliation{Thomas  Jefferson National Accelerator Facility, Newport News, VA 23606, USA}  
\affiliation{Physics Department, College of William and Mary, Williamsburg, VA 23187, USA}
\date{July 8, 2005}
\begin{abstract}
We discuss the links between Generalized Parton Distributions (GPDs) 
and elastic nucleon form factors. These links, in the form of sum rules, 
represent powerful constraints on parametrizations of GPDs.
A Regge parametrization for  GPDs at small momentum transfer, 
is extended to the large momentum transfer region and it is found to 
describe the basic features of proton 
and neutron electromagnetic form factor data. 
This parametrization is used to estimate the quark contribution to the 
nucleon spin. 
\end{abstract}
\pacs{12.38.Bx, 13.60.Hb,  13.60.Fz, 13.60.Le }
\maketitle

\newpage

\section{Introduction}
Generalized parton distributions (GPDs)
\cite{Muller:1998fv,Ji:1996ek,Radyushkin:1997ki}  
are universal non-perturbative objects entering the
description of hard exclusive electroproduction processes 
(see Refs.~\cite{Ji:1998pc,Radyushkin:2000uy,Goeke:2001tz,Diehl:2003ny,Belitsky:2005qn}
for reviews and references).
These GPDs, which are defined for each quark flavor ($u$, $d$, $s$), 
parametrize nonforward matrix 
elements of lightcone operators.
They depend upon the  longitudinal momentum fractions 
of the initial and final quarks  and upon the
overall momentum transfer $t$ to the nucleon. 
When the momentum fractions $x+\xi,x-\xi$ of initial and final quarks are
different ($\xi$ being the longitudinal momentum
asymmetry, or skewness), one accesses quark momentum correlations in the
nucleon. Furthermore, if one of the quark momentum fractions is
negative, GPDs reflect  an antiquark contribution,
and consequently one can 
investigate $q \bar q$ configurations in the nucleon. 
Therefore, these functions contain a wealth of new nucleon 
structure information, generalizing  that obtained from 
inclusive deep inelastic scattering.  
\newline
\indent
In hard exclusive processes, such as deeply virtual Compton scattering, 
GPDs enter in most observables through convolution integrals. 
Hence, to access  GPDs, the most 
realistic strategy to date seems through judicial parametrizations.
 Building  self-consistent  models of GPDs is, however,  a rather 
 difficult problem, because one needs to satisfy many physical principles  and 
 constraints which should be obeyed by GPDs. They include   
  spectral properties,
  polynomiality condition, positivity, relations
 to parton densities and form factors 
 \cite{Muller:1998fv,Ji:1996ek,Radyushkin:1997ki,Ji:1998pc}.
\newline
\indent
In this paper, we elaborate on the $t$-dependence of the $\xi=0$ 
 generalized parton distributions, 
and its interplay with the $x$-dependence. This subject   has 
attracted a considerable interest.
In particular, it has been shown \cite{Burkardt:2000za,Burkardt:2002hr,Diehl:2002he}
that by a Fourier transform of the
$t$-dependence of GPDs, it is conceivable to access the spatial 
distribution of partons in the transverse plane, 
and to provide a 3-dimensional picture of the 
nucleon~\cite{Ralston:2001xs,Belitsky:2003nz}.  
The $t$-dependence of moments of GPDs has also become amenable to 
lattice QCD calculations~\cite{:2003is} recently. As the lattice calculations 
mature further, they may eventually provide additional constraints on 
moments of generalized parton distributions. 
Phenomenological estimates of the $t$-dependence 
and $t$-dependent parametrizations of GPDs have 
 already been discussed in Refs. 
\cite{Radyushkin:1998rt,Diehl:1998kh,Afanasev:1999at,Stoler:2001xa,Stoler:2003mx,Burkardt:2002hr,Belitsky:2003nz},
and more recently, in 
 Ref.~\cite{Diehl:2004cx}. 
 Some results of the present paper were reported in 
 Refs.~\cite{Vanderhaeghen:2002pg,Radyushkin:2004sr}.
\newline
\indent
We give here several parametrizations of the $t$-dependence of the GPDs, 
both at small and large values of $-t$ (with $ t < 0$, i.e. in the 
spacelike region).  
We start in Section~\ref{sec:sumrule} by reviewing the relevant sum 
rules which link GPDs to form factors. 
Subsequently, we discuss in Section~\ref{sec:gaussian} a Gaussian ansatz 
for the $t$-dependence of GPDs (at large $-t$) 
which has been introduced and used in Refs. 
\cite{Radyushkin:1998rt,Diehl:1998kh}. Such a Gaussian ansatz, 
however, is not able to describe the small $-t$ behavior of GPDs,  and in 
particular gives divergent rms radii for the nucleon electromagnetic form 
factors. We therefore proceed in Section~\ref{sec:regge1} to describe  a 
Regge parametrization 
\cite{Goeke:2001tz,Vanderhaeghen:2002pg}
which provides a physically consistent behavior of 
form factors at small $-t$. We extend this model then in 
Section~\ref{sec:regge2} to large $-t$ so as to yield the observed 
power behavior of the electromagnetic form factors at large 
(spacelike) momentum transfers. 
We found   a quite economical parametrization that  allows for a 
description of both proton and neutron electromagnetic form factors
with only  3  parameters: the universal Regge slope $\alpha^{\prime}$
and two parameters $\eta_u,\eta_d$ governing the $x \to 1$
behavior of the splin-flip GPDs
$ {\cal E}^{u} (x,t=0), {\cal E}^{d} (x,t=0)$ relative to that of 
the usual  parton densities $ {u} (x),{d} (x)$. 
 We discuss the comparison of our results  with the data 
in Section~\ref{sec:results}, and use our parametrization
  to estimate the 
quark contribution to the nucleon spin.
In Section~\ref{sec:impact}, we  discuss the positivity constraints 
on  GPDs  in the impact parameter ${\bf b_\perp}$ representation. 
To extend the region in  $x$ and  ${\bf b_\perp}$
where the positivity constraints are satisfied, we   
 propose  a  model 
in which  the parameters $\eta_u$ and $\eta_d$
are equal. It provides (with just two parameters) almost the same  
quality description 
of the four form factors as the 3-parameter model. 
Our conclusions are presented in 
Section~\ref{sec:conclusions}.

\section{Form factors and GPDs}
\label{sec:sumrule}

The nucleon Dirac and Pauli form factors $F_1(t)$ and $F_2(t)$ 
\begin{eqnarray}
F_i(t) = \sum_q e_q F_i^q (t)
\label{eq:Fi} 
\end{eqnarray}
can be calculated from the valence quark GPDs $H$ and $E$ through the 
following sum rules for their  flavor components ($q = u, d$) 
\begin{eqnarray}
&& F_{1}^{q}(t) \,=\, \int_{-1}^{+1}dx\; H^{q}(x,\xi ,t) \,  ,
\label{eq:hsumrule} \\
&& F_{2}^{q}(t) \,=\, \int_{-1}^{+1}dx\; E^{q}(x,\xi ,t) \, \; .
\label{eq:esumrule} 
\end{eqnarray}
Since the result of the integration does not depend 
on the skewness $\xi$,  one can choose $\xi = 0$ in the previous equations. 
Furthermore, the integration region can be reduced to the $0 < x <1$
interval, introducing the nonforward parton densities~\cite{Radyushkin:1998rt}:
\begin{eqnarray}
{\cal H}^q (x,t) &\,=\,& H^q(x,0 ,t) + H^q(-x, 0, t)    , \\ 
{\cal E}^q (x,t) &\,=\,& E^q(x,0 ,t) + E^q(-x, 0, t) ,
\end{eqnarray}
obeying  the conditions 
\begin{eqnarray}
\int_{0}^{1}dx\; {\cal H}^q (x,t) = F_{1}^{q}(t) \  , 
\label{eq:F1q} \\ 
\int_{0}^{1}dx\; {\cal E}^q (x,t) = F_{2}^{q}(t) \  , 
\label{eq:F2q}  
\end{eqnarray}
that follow from  the sum rules (\ref{eq:hsumrule}), (\ref{eq:esumrule}). 
The  ${\cal H}^q (x,t)$ functions also satisfy  the $t \to 0$ reduction relations
 \begin{eqnarray}
 {\cal H}^u (x,t=0)  = u_v  (x),  \qquad  
\qquad  {\cal H}^d (x,t=0) = d_v  (x), 
\end{eqnarray}
connecting  them with the usual valence quark densities in the proton. 
The $t=0$ limit of the ${\cal E}^q (x,t)$  distributions exists,
but  the ``magnetic'' densities ${\cal E}^q (x,0) \equiv {\cal E}^q (x)$  
cannot be directly expressed
in terms of any  known parton distribution: they contain 
new information about the nucleon structure.
However, the normalization integrals
 \begin{eqnarray}
 \kappa_q \equiv \int_{0}^{1}dx\; {\cal E}^q (x) 
 \label{eq:kappas} 
\end{eqnarray}
are constrained by the requirement that the values 
$F_2^{p}  (t=0)$ and $F_2^{n}  (t=0)$ are equal to  the anomalous magnetic
moments of the proton and neutron.  
This gives
\begin{eqnarray}
\kappa_u \,&=&\, 2 \, \kappa_p \;+\; \kappa_n \approx +1.673 \; , 
\label{eq:kappau} \\
\kappa_d \,&=&\, \kappa_p \;+\; 2 \, \kappa_n \approx  -2.033 \; .
\label{eq:kappad}
\end{eqnarray}
For comparison, the normalization 
integrals for the ${\cal H}^u (x)=u_v(x)$ and 
${\cal H}^d (x)=d_v(x)$ distributions are given by 2 and 1 
respectively, the 
number of $u$ and  $d$ valence quarks in the proton. 

\section{Gaussian ansatz}
\label{sec:gaussian}

The simplest  model for the proton's ${\cal H}^q (x,t)$  is 
to separate the $x$ and $t$-dependencies and express it as the 
product 
\begin{eqnarray}
{\cal H}^q (x,t) = q_v(x) F_1(t)
\end{eqnarray}
of the parton density
$q_v(x)$   and the  $F_1(t)$ form factor of the proton.
It trivially reproduces $q_v(x)$ in the forward limit
and gives the correct result for $F_1^p(t)$. 
However, such a complete factorization of the $x$ and $t$ 
dependencies seems rather unrealistic. 
In particular, the form factor formula \cite{Drell:1969km}
\begin{eqnarray}
&&F(q_{\perp}^2) = \nonumber \\ 
&&\sum_{n=1}^{\infty} \int  \prod_{i=1}^n d^2k_{i_{\perp}}
d x_i 
\sum_a e_a \Psi^*_{P^{'}} (x_1, 
\ldots , x_n; k_{1_{\perp}}- x_1 q_{\perp}, \ldots , k_{a_{\perp}} 
+ (1-x_a) q_{\perp} , \ldots ,k_{n_{\perp}}  - x_n q_{\perp}) 
\nonumber \\ &&\hspace{3cm}\times
\Psi_{P} (x_1, \ldots , x_n; k_{1_{\perp}}, \ldots , k_{a_{\perp}} , 
\ldots ,k_{n_{\perp}} ) \, \delta^{(2)}  \left (\sum_{i=1}^n k_{i_{\perp}} 
\right ) 
\, \theta \left (1- \sum_{i=1}^n x_i \right ), 
\label{eq:drellyan}
 \end{eqnarray}
of the light-cone formalism is a  
convolution  of the light cone wave functions containing 
nonfactorizable combinations $k_{i_{\perp}}  - x_i q_{\perp}$.
Furthermore, the $n$-body Fock component
 $\Psi_{P} (x_1, \ldots , x_n; k_{1_{\perp }}, \ldots  
,k_{n_{\perp}} ) $ 
of the light-cone wave function 
  usually  depends   on the transverse momenta $\{ k_{i_{\perp }}\}$
through the 
$\sum_i k_{i_{\perp}}^2/x_i$ combination
involving both $ k_{i_{\perp}}$ and  the fractions $x_i$ of the
hadron longitudinal momentum carried by the quarks. 
 If the dependence on this combination has a Gaussian form, 
 the $k_{\perp}$ integration can be performed analytically
 providing  an example of the interplay between the $x$ and $t$ dependencies.
 The result of  integration can be most easily 
 illustrated on the simplest example of a two parton system ($n = 2$). 
 In this case 
\begin{eqnarray}
F^{(2)} (q_{\perp}^2) = \int_0^1 dx \, d^2k_{\perp} 
 \Psi^* (x; k_{\perp } +(1-x)q_{\perp} )  \Psi (x; k_{\perp }) \ .
 \label{eq:dytwobody}
 \end{eqnarray}
 Assuming the Gaussian ansatz  
 \begin{eqnarray} 
 \Psi (x; k_{\perp }) \sim \exp \left [ -\frac{k_{\perp }^2}{2x\lambda^2} - 
 \frac{k_{\perp }^2}{2(1-x) \lambda^2}\right ] = 
 \exp \left [ -\frac{k_{\perp }^2}{2x(1-x)\lambda^2} \right ] \ , 
 \end{eqnarray}
we obtain 
\begin{eqnarray}
F^{(2)}(q_{\perp}^2) = \int_0^1 dx \, q^{(2)} (x)  e^{-(1-x)q^2_{\perp}/4x\lambda^2} \ ,
\end{eqnarray}
where $q^{(2)} (x)$ has the meaning of the two-body part of the quark 
density  $q(x)$. 
This suggests the Gaussian (G) parametrization 
\cite{Barone:ej,Radyushkin:1998rt} for the nonforward parton densities 
\begin{eqnarray}
 {\cal H}^q_G (x,t) = q_v(x) \, e^{(1-x)t/4x\lambda^2} ,
 \label{eq:gauss}
 \end{eqnarray}
containing a nontrivial interplay
 between $x$ and $t$ dependencies.
  The scale $\lambda^2$ characterizes 
 the average transverse momentum of the valence quarks
 in the nucleon. 
 The best agreement (within 10\%) between 
 experimental data for   $F_1^p(t)$ in the 
 moderately large $t$ region 1\,GeV$^2 < -t <10$\,GeV$^2$ 
 and calculations based on 
Eqs.~(\ref{eq:Fi}), (\ref{eq:F1q}), (\ref{eq:gauss}) is obtained 
 for 
 $\lambda^2 \sim 0.7\,$GeV$^2$. This value 
 corresponds 
 to an average transverse momentum of about 300 MeV \cite{Radyushkin:1998rt},
 which  is close to the inverse of the proton size.  
 The latter can also be estimated by calculating  the 
 mean squared radius 
\begin{eqnarray}
 r^2_{1, p} \,=\, 6 \left. \frac{dF_1^p(t)}{dt} \right |_{t=0} \ .
\end{eqnarray}
 The Gaussian model for ${\cal H}^q (x,t)$ then 
 gives the expression 
\begin{eqnarray}
 r^2_{1, p} \,=\, 6 \int _{0}^{1}dx \; 
\biggl\{ e_u \, \, u_v(x) \,+\, e_d \, \, d_v(x) \biggr\} \, \frac{1-x}{x} \; .
\label{eq:gaussrms1p}
\end{eqnarray}
If one assumes
the standard Regge-type behavior  $q_v(x)|_{x\to 0} 
 \sim x^{-0.5}$ of the parton densities at small $x$,
 the integral in (\ref{eq:gaussrms1p}) diverges. 
 To get a finite slope we should   modify  the model for ${\cal H}^q (x,t)$ 
 in the region of small $x$.

\section{Small t behavior and Regge parametrization (R1)}
\label{sec:regge1}

The Regge picture suggests 
a $x^{-\alpha (t)}$ behavior at small $x$ or the   
\begin{eqnarray}
{\cal H}^q (x,t) = q_v(x)  x^{-(\alpha (t)- \alpha (0))}  
\end{eqnarray}
model for the nonforward densities 
${\cal H}^q (x,t)$.  Assuming a linear  Regge trajectory 
with the slope  $\alpha^{\, '}$, we get  
\begin{eqnarray}
{\cal H}^q_{R1} (x,t) = q_v(x) \  x^{- \alpha^{\, '} \, t}  \  .
\label{eq:hr1}
\end{eqnarray}
This ansatz  was already discussed  in Ref.~\cite{Goeke:2001tz}.
The  $u$ and $d$ flavor components of the Dirac form factor
are then given by 
\begin{eqnarray}
F_1^u(t) \,=\, \int _{0}^{1}dx \; u_v(x) \;
{e^{-t\, \alpha^{\, '}  \ln x} } \qquad , \qquad
F_1^d(t) \,=\, \int _{0}^{1}dx \; d_v(x) \; 
{e^{-t \, \alpha^{\, '}  \ln x}}  \  . 
\label{eq:f1_1}
\end{eqnarray}
The proton and neutron Dirac form factors  follow from 
\begin{eqnarray}
F_1^p(t) \,=\, e_u \, F_1^u(t) \;+\; e_d \, F_1^d(t) \, , 
\label{eq:f1p} \\
F_1^n(t) \,=\, e_u \, F_1^d(t) \;+\; e_d \, F_1^u(t) \, . 
\label{eq:f1n} 
\end{eqnarray}
By construction $F_1^p(0)$ = 1, and $F_1^n(0)$ = 0.
The Dirac mean squared radii of proton and neutron 
in this model are given by
\begin{eqnarray}
r^2_{1, p} &\,=\,& -6 \, \alpha^{\, '} \,
\int _{0}^{1}dx \; 
\biggl\{ e_u \, \, u_v(x) \,+\, e_d \, \, d_v(x) \biggr\} \, \ln x \; ,
\label{eq:rms1p} \\
r^2_{1, n} &\,=\,& -6 \, \alpha^{\, '} \,
\int _{0}^{1}dx \; 
\biggl\{ e_u \, \, d_v(x) \,+\, e_d \, \, u_v(x) \biggr\} \, \ln x  \;.
\label{eq:rms1n}
\end{eqnarray}
Instead of the $1/x$ factor present in the Gaussian model,
we have now a   much softer logarithmic singularity 
at small $x$, and the integrals for  $r^2_{1}$ converge.  
\newline
\indent
To calculate $F_2$, we need  an ansatz for the  nonforward
parton densities ${\cal E}^q (x,t)$. We assume the same 
Regge-type structure  
\begin{eqnarray}
{\cal E}^q_{R1}  (x,t) = {\cal E}^q (x)\,  x^{-\alpha^{\, '} t}
\end{eqnarray}
as for ${\cal H}^q (x,t)$. 
The next step is 
to model the forward magnetic densities  ${\cal E}^q (x)$.
The simplest idea is to take them proportional 
to the ${\cal H}^q (x)$  densities.
Choosing 
\begin{eqnarray}
{\cal E}^u (x) = \frac{\kappa_u}{2} u_v(x)  
 \qquad {\rm and}  \qquad {\cal E}^d (x) = \kappa_d d_v(x) \  ,
\label{eq:er1}
\end{eqnarray}
we  satisfy the normalization conditions
(\ref{eq:kappas}) which, in their turn, guarantee that 
$F_2^p(0)$ = $\kappa_p$, and $F_2^n(0)$ = $\kappa_n$.
\newline
\indent
As we will show in Section~\ref{sec:results}, 
the Regge model R1 fits 
$F_1^p(t)$ and $F_2^p(t)$ data for small momentum transfers 
$-t \lesssim 0.5$\,GeV$^2$.
However,  the  suppression at larger $-t$ in the R1 model is too strong, 
and it 
consequently falls considerably short of the data for $-t > 1$\,GeV$^2$.

\section{Large t behavior and modified Regge parametrization (R2)}
\label{sec:regge2}

To improve the agreement with the data at large $-t$,
we need to modify our models. 
Note, that both the Gaussian (G) and the Regge-type
   model (R1)  discussed above have the structure 
$${\cal H}(x,t) = 
q_v(x) \exp[\,tg(x)] \ ,
$$ 
with $g(x) \sim (1-x)/x$ and 
$g(x) \sim -\ln x$, respectively. Hence, at large $t$, 
the form factors are  dominated by 
integration over  regions where 
$tg(x) \sim 1$ or $g(x) \sim 1/t \to 0$.
In both cases, $g(x)$ vanishes only for 
$x \to 1$, and  the large-$t$ asymptotics 
of $F_i(t)$ is governed by the 
$x \to 1$ region. Given 
$g(x)\sim 1-x$ as  $x \to 1$, one derives  that
if  $q_v(x)\sim (1-x)^{\nu}$ for $x$ close to 1, then 
the form factors drop like $1/t^{\nu +1}$ at large $t$.
Experimentally,   $\nu$ is close to 3, 
thus the models G and R1  correspond to
 the $\sim 1/t^4$  behavior  
for the form factors. This seems to be in contradiction with 
the experimentally established $1/t^2$ 
behavior of $F_1^p (t)$,
so one may be tempted to  conclude 
that these models have no  chance to describe  the data. 
A trivial but important  remark is that 
the model curves for $F_1^p (t)$ are more complicated functions
than  just a pure 
power behavior $\sim~1/t^4$.  
In fact, up to 10\,GeV$^2$,  the Gaussian model  reproduces 
the data for $F_1^p$ within 10\%~\cite{Radyushkin:1998rt}.
For higher $t$,   the Gaussian model prediction for $F_1^p$  
drops faster than $1/t^2$ and goes below the data.
 However, the nominal  $1/t^4$ asymptotics 
is achieved only at very large values  $-t\sim$ 500\,GeV$^2$.
As we show in Section~\ref{sec:results},  
the Regge-type model R1 result 
visibly underestimates the data for $F_1^p$ already for 
$-t\sim$ 1\,GeV$^2$  though   one should wait till $-t\sim$ 100\,GeV$^2$
to see that the $1/t^4$ behavior really settles.   
 Thus, the conclusions made 
on   the basis of asymptotic relations
might  be of little importance in the experimentally
accessible  region: a curve with a ``wrong'' 
 large-$t$ behaviour might be quite successful phenomenologically
 in a rather wide range of $t$.
\newline
\indent
The shortcomings of the G and R1  models are more of a theoretical
nature. Namely, they do not satisfy the Drell-Yan (DY)
relation \cite{Drell:1969km,West:1970av}
  between the $x \to 1$ behavior 
of the structure functions  and the 
$t$-dependence of elastic form factors. According to  DY, 
 if the parton density
behaves like $(1-x)^{\nu}$, then the relevant 
form factor should decrease as $1/t^{(\nu+1)/2}$ for large $t$.
Such a relation does not  hold if $g(x) \sim 1-x$ 
but it holds if $g(x) \sim (1-x)^2$.
Thus, the simplest idea is to attach  an extra $(1-x)$ factor 
to the original $g(x)$ functions.   
To preserve  the Regge structure at small
$x$ and $t$ we take the modified Regge ansatz R2 
\cite{Burkardt:2002hr,Burkardt:2004bv} 
\begin{eqnarray}
{\cal H}^q_{R2} (x,t) = q_v(x)  x^{-\alpha^{\, '} (1-x)t}  \  .
\label{eq:hr2}
\end{eqnarray}
\indent
 The inability of the G parametrization  to satisfy the 
 DY relation  may seem rather     surprizing 
in view of the fact that the original  derivation of the relation 
by Drell and Yan \cite{Drell:1969km} 
is based on the analysis of the 
large-$q_{\perp}$  limit of the general formula (\ref{eq:drellyan}) 
of which the G ansatz is a specific case corresponding 
to $n=2$ and the 
$\Psi (x; k_{\perp }) \sim \exp [ -{k_{\perp }^2}/{2x(1-x)\lambda^2}  ]$
wave function. 
Note, that if  the wave function $\Psi (x,k_{\perp} )$ depends on $k_{\perp}$ 
 through the combination $k_{\perp}^2/x+k_{\perp}^2/(1-x)$,
 then the restriction on  the $x\to 1$ integration 
 region should be $|k_{{\perp}} 
+ (1-x) q_{\perp}|^2/(1-x) \lesssim \lambda^2 $ which results in   the 
$1-x \lesssim \lambda^2/q^2_{\perp }$ constraint on the $x$ integration. 
Also, 
from the explicit form of the Gaussian parametrization (\ref{eq:gauss}), 
 it is clear that the essential region for the $x_a$ integration 
 is $1-x_a \sim \lambda^2/(-t)$ which gives the $1/t^{\nu+1}$
 result, that differs from the canonical $1/t^{(\nu+1)/2}$  DY 
 prediction. 
The resolution of this discrepancy is rather simple.
In fact, in the derivation given by Drell and Yan,
 it was implied that the wave function depends on $k_{\perp}$
 through the combination $(k_{\perp}^2+m_q^2)/x(1-x)$,
 with $m_q$ being 
the (constituent) quark mass.
 Then, in the Gaussian case, after the $k_{\perp}$-integration,
 one would have the structure 
 $\sim \exp \{   -[(1-x)q_{\perp}^2+m_q^2/(1-x)]/\lambda^2] \}$
in the $x \sim 1$ region, and
at  large $q_{\perp}^2$ the dominant contribution comes from 
the region $1-x \sim m_q/q_{\perp}$.
This agrees  with the argumentation of Ref.  \cite{Drell:1969km}, 
 that  the leading contribution 
to the form factor is due to  integration over 
the region $1-x_a < m_q/q_{\perp }$ 
where the longitudinal momentum fraction $x_a$ of the active quark
is close to 1 and those of the passive quarks are close to 0,
so that $  |k_{a_{\perp}} 
+ (1-x_a) q_{\perp}|$  and  all $|k_{i_{\perp}}  - x_i q_{\perp}|$ 
are bounded by $O(\lambda)$. Integration over all $k_{i_{\perp}}$'s 
and $x_i$'s of  passive quarks gives $q(x_a)$.
If $q(x_a) \sim (1-x_a)^{\nu}$, then the final integration over the region  
$x_a\sim 1-\lambda/q_{\perp }$ gives
$F(q_{\perp}) \sim 1/q_{\perp}^{\nu +1}\sim 1/t^{(\nu+1)/2}$.
\newline
\indent
Turning back to the
Gaussian model with zero quark mass,
it is  easy to realize that the factor 
$(1-x)t/x\lambda^2$ in the exponent of the
G parametrization may  be viewed as $[(1-x)q_{\perp}]^2/ x(1-x)\lambda^2$ 
with  $1/x(1-x)$ coming from the $\exp[-k_{\perp}^2/2x(1-x)\lambda^2]$ 
structure of the $k_{\perp}$-dependence of the wave function.
As we have seen, to get the Regge-type behavior at small $x$,
one should soften  the $1/x$ factor in the exponential 
substituting it by  $\ln \, x$. Since  the limit $x_a \to 1$ 
for the active quark corresponds to the Regge 
limit $x_s \to 0$ for the spectators,  
one may  expect by analogy that  the $1/(1-x)$  singularity
is also softened after inclusion of higher Fock components.
The R2 ansatz  corresponds to
substitution of the $1/(1-x)$ factor by a constant.
Other arguments in favor of the
R2 model can be found in 
Ref. \cite{Burkardt:2004bv}.

The  correlation between  the power behavior of form factors and  the
behavior of inclusive structure functions $W (x_{\rm B})$ of deeply inelastic
scattering at large Bjorken variable $x_{\rm B}$ is a rather 
popular subject (``inclusive-exclusive connection'').
The basic idea behind the possibility of such a correlation 
is that, for sufficiently large $x_B$, one approaches the
exclusive single-hadron pole. The invariant mass $W^2 = (p + q)^2$ of the
 hadronic system produced in deep inelastic scattering is related
to the Bjorken variable $x_{\rm B}$ by
\begin{equation}
\label{XbMass}
1 - x_{\rm B} = x_{\rm B} \, \frac{W^2 - M_h^2}{{ Q}^2}
\, ,
\end{equation}
 and the single-hadron
contribution to cross section is given by the form factor squared multiplied by
$\delta (W^2 - m_h^2)$. The  Bloom-Gilman duality idea \cite{Bloom:1970xb}
is that  the
$W^2$-integral of the hadron contribution is equal to the $x$-integral of the
structure function $W_1 (x)$ over a duality region with fixed boundaries in
the variable $W^2$. This gives a relation between the power $\nu$ specifying
the $(1 - x)^\nu$ behavior of the structure function $W_1 (x)$ in the $x \to 1$
region and the power-law behavior of the squared elastic form factor: $F^2
(t) \sim (1/|t|)^{\nu + 1}$. In the proton case, with
usually adopted value 
 $\nu = 3$, one obtains  a dipole behavior for the Dirac $F_1 (t)$ form
factor.

We would like to strongly  emphasize here
that one should not confuse the Bloom-Gilman duality with the
Drell-Yan relation \cite{Drell:1969km}. As we discussed above,
the latter connects some   integral of a nonforward
parton density ${\cal H}^q (x, t)$ over the interval $x > 1- \lambda/{ \sqrt{-t}}$
 with the  first power of the form
factor. It is worth  to repeat and stress the statement:
the Bloom-Gilman relation connects 
an  $x$ integral of the structure function 
with the {\it square} of the form factor,
while the DY relation expresses an(other)  $x$ integral of the structure function 
in terms of the {\em first power} of the form factor. 
Moreover, the dominance of the region $x > 1- \lambda/\sqrt{-t}$ 
implied by the DY relation is a consequence
of a specific structure of  the density  ${\cal H}^q (x, t)$,
the interplay between its $x$ and $t$ dependence. As we have seen, 
the Drell-Yan relation does not work for the Gaussian model, but it 
holds for the modified  Regge model R2.

One should also realize that both relations were formulated before the QCD era,
and  in  absolutely nonperturbative
terms. Their authors  did not   assume that the
 shape of the structure function $F_1 (x)$ or that of the nonforward parton densities
${\cal H}^q (x,t)$ are   generated by perturbative QCD dynamics based on  hard gluon exchanges. 
Their prescription was that knowing the $x \to 1$ behavior of the 
structure functions, one can use Bloom-Gilman or Drell-Yan
relations to get predictions for form factors. Both relations have a common feature:
if one changes  the power $\nu$ in the $(1 - x)^\nu$ behavior of the structure function,
this would result 
 in a change of the $1/( -t)^{(\nu+1)/2}$ power behavior of the form factor, i.e., 
 the powers themselves are not fixed, 
 what is fixed is the relation(s) between them. 
Accidentally, both relations give the same correlation 
between the two powers, and that is why
they are  confused sometimes.

In distinction to the Bloom-Gilman and Drell-Yan relations, 
perturbative QCD  predicts definite powers
for the asymptotic  behavior of form
factors and the $x \to 1$ behavior of parton distributions. For example, it gives  
$(\alpha_s/|t|)^{n-1}$ for a spin-averaged form factor of an 
$n$-quark hadron, and  it also predicts fixed powers $\alpha_s^{2n - 2} (1-x)^{2n - 3}$
for the $x \to 1$ behavior of its valence quark distributions 
(see \cite{Lepage:1980fj}) \footnote{We would like to comment here 
that all existing phenomenological 
parametrizations
of parton densities based on fits to data 
 ascribe a larger  power of $(1-x)$ for the  $d$ quark 
distribution compared to the $u$ one, uniformly accepting a   \ 
 $\sim (1-x)^4$ behavior rather than pQCD's $(1-x)^3$ form for  $d$  quarks.}. 
The basic difference between the pQCD formulas and BG \& DY relations
is that a particular power behavior of a hadronic form factor in pQCD 
{\em is not a consequence} of  a particular limiting power behavior
of the respective parton distribution in the region $x \to 1$.
The fixed powers predicted by   pQCD are correlated 
simply because of similarity of the relevant diagrams, but 
there is no causal connection between them. 
Also, though the powers predicted by pQCD for 
the nucleon are in agreement with   BG and DY relations,
it was never  demonstrated that 
there is  a fundamental reason behind  this fact.

Formally, the relevant powers of
$(1 - x)$, $1/|t|$ and $\alpha_s$ for the proton 
are correlated in pQCD just like  in the Bloom-Gilman
relation. However, a direct calculation of pQCD diagrams for $W_1 (x)$ gives
expressions which have more complicated structure than the squares of form factors
(see e.g., \cite{Yuan:2003fs}, where the $x \to 1$ behavior of 
GPDs is also discussed). Thus, it is not clear yet  
if the Bloom-Gilman relation works in pQCD. 

With the Drell-Yan relation, the situation is simpler.
The whole logic of the hard-rescattering pQCD mechanism  
is orthogonal 
to the Feynman-Drell-Yan approach. 
In the pioneering paper by Lepage and Brodsky \cite{Lepage:1980fj},
it was emphasized in the Introduction of that  paper that the Drell-Yan relation
is invalid in pQCD.  
It was stressed, in particular, that  the correlation  between the  
powers of  $\alpha_s$ in  pQCD predictions 
disagrees with the Drell-Yan relation. 
For instance,  the leading $(1 - x)^3$ term in $W_1 (x)$ for the nucleon is attributed 
in pQCD to diagrams involving four  hard gluon exchanges, and is accompanied hence  by the
$\alpha_s^4$ factor. Integrating it over the region $x> 1- \lambda/{ Q}$, one
would get a contribution $\sim \alpha_s^4/{t}^2$ that has the same $1/{ t}^
2$ power as the pQCD prediction for the nucleon form factor, but has two extra 
 powers of $\alpha_s$.

Our models imply  the  dominating role of the Feynman-Drell-Yan mechanism
for the hadronic form factors, and we assume that 
 the $x \to 1$ behavior of the parton distributions is  generated 
 by nonperturbative dynamics. 
In this scenario,  the observed behaviour of  hadronic form factors
is also due to the  nonperturbative dynamics, and 
 we treat as negligible the pQCD contributions to 
the nucleon form factors, 
which have  $(\alpha_s/\pi)^2$ suppression compared to
the nonperturbative terms.  
\newline
\indent
In the following estimates we take the unpolarized parton distributions 
at input scale $\mu^2$ = 1 GeV$^2$ from the 
MRST2002 global NNLO fit~\cite{Martin:2002dr} as~:  
\begin{eqnarray}
u_v &=& 0.262 \, x^{-0.69} (1 - x)^{3.50} 
\left( 1 + 3.83 \, x^{0.5} + 37.65 \, x \right),  \\
d_v &=& 0.061 \, x^{-0.65} (1 - x)^{4.03} 
\left( 1 + 49.05 \, x^{0.5} + 8.65 \, x \right) . 
\end{eqnarray}
One sees that $\nu_u = 3.50$ and $\nu_d = 4.03$ at a scale $\mu^2$ = 1~GeV$^2$.
Hence,  the asymptotic behavior of $F_1^p (t)$
in the R2 model is $1/t^{2.25}$, generating a slightly
faster decrease than the ``canonical'' $1/t^2$.
Again, this asymptotic limit sets in for very large $t$ 
values.
\newline
\indent
At small $t$, the modifications
compared to the R1 model 
are not very significant numerically.
The Dirac mean squared radii of proton and neutron in the R2 model are finite and given 
by
\begin{eqnarray}
r^2_{1, p} &\,=\,& - 6 \, \alpha^{\, '} \, 
\int _{0}^{1}dx \; 
\biggl\{ e_u \, \, u_v(x) \,+\, e_d \, \, d_v(x) \biggr\} \, (1 - x) \,\ln x  
\; , \label{eq:rms1p_2} \\
r^2_{1, n} &\,=\,& - 6 \, \alpha^{\, '} \, 
\int _{0}^{1}dx \; 
\biggl\{ e_u \, \, d_v(x) \,+\, e_d \, \, u_v(x) \biggr\} \, (1 - x) \, \ln x  
\; . \label{eq:rms1n_2}
\end{eqnarray}
\indent
In case of   the Pauli form factor $F_2$, we perform  the same modification 
of the  ansatz 
for the ${\cal E}^q (x,t)$ densities taking 
\begin{eqnarray}
{\cal E}^q (x,t) = {\cal E}^q (x)  x^{-\alpha^{\, '} (1-x)t}  \  .
\label{eq:er2}
\end{eqnarray}
\indent
Experimentally,  the proton helicity flip  form factor $F_2(t)$ has a faster 
power fall-off  
at large $t$ than $F_1(t)$.  Within all our  models, 
this  means that the  $x\sim 1$ behavior of
the functions ${\cal E}(x)$ and ${\cal H}(x)$ should be different.
To produce a faster decrease  with $t$, the $x\sim 1 $ limit of the density 
 ${\cal E}^q(x)$ should have extra powers of $1-x$ compared to that  of 
 ${\cal H}^q(x)$ (in case of the G model, such a modeling was originally incorporated 
 in Ref. \cite{Afanasev:1999at}). 
 Aiming to avoid introducing too many free
parameters, we try the simplest ansatz for ${\cal E}^q(x)$ in which we 
get them by just multiplying the 
valence quark distributions  by an additional
factor $(1 - x)^{\eta_q}$, i.e., we take 
\begin{eqnarray}
{\cal E}^u (x) = \frac{\kappa_u}{N_u} (1-x)^{\eta_u} u_v(x)  
 \qquad {\rm and}  \qquad {\cal E}^d (x) = \frac{\kappa_d}{N_d} 
 (1-x)^{\eta_d} d_v(x) \  ,
\label{eq:e2}
\end{eqnarray}
where  the normalization factors $N_u$ and $N_d$  
\begin{eqnarray}
N_u &\,=\,&  \int _{0}^{1}dx \; (1 - x)^{\eta_u} \, u_v(x) \; , 
\label{eq:nu} \\ 
N_d &\,=\,&  \int _{0}^{1}dx \; (1 - x)^{\eta_d} \, d_v(x) \; 
\label{eq:nd}  
\end{eqnarray} 
guarantee the  conditions (\ref{eq:kappas}). 
The flavor components of the 
Pauli form factors are now given  by 
\begin{eqnarray}
F_2^u(t) \,&=&\, \int _{0}^{1}dx \, 
 {\kappa_u  \over {N_u}} \, (1 - x)^{\eta_u} \, u_v(x) \, 
{{x^{-\alpha^{\, '} \, (1 - x) \, t}}} \;, 
\label{eq:f2u_2}\\
F_2^d(t) \,&=&\, \int _{0}^{1}dx  
\, {\kappa_d  \over {N_d}} \, (1 - x)^{\eta_d} \, d_v(x) 
{{x^{-\alpha^{\, '} \, (1 - x) \, t}}} \; .
\label{eq:f2d_2}
\end{eqnarray}
\indent
The powers $\eta_u$ and $\eta_d$ are
 to be determined from a fit 
to the nucleon form factor data.  Note that the value $\eta_q = 2$ 
corresponds to a $ 1/t$ asymptotic behavior of the ratio
$F_2^q(t)/F_1^q(t)$ at large $t$. 
We also tried an even simpler 2-parameter version of the R2 model, with 
$\eta_u, \eta_d$ restricted to be equal to each  other $\eta_u = \eta_d$.

\section{Results} 
\label{sec:results}

In this section, we show the results for the proton and neutron 
electric and magnetic form factors based on the Regge and modified 
Regge parametrizations discussed in this work. 
In recent years, a lot of high accuracy data have become available 
for the nucleon electromagnetic form factors 
in the spacelike region, which put stringent 
constraints on our parametrizations of GPDs. 
\newline
\indent
The parametrization R1 of Eqs.~(\ref{eq:hr1},\ref{eq:er1}) depends on 
only one parameter~: $\alpha^{\prime}$,  
which can only be varied within a narrow range if it is  to be 
interpreted as a slope of the Regge trajectory. The modified 
Regge parametrization R2 of Eqs.~(\ref{eq:hr2},\ref{eq:er2}) depends on 
three  parameters. Besides $\alpha^{\prime}$, 
it also depends on $\eta_u$ and $\eta_d$, which govern the $x \to 1$ behavior 
of the GPD $E$, that in turn is determined from the 
behavior of $F_2^p / F_1^p$ at large $-t$.  
In determining these parameters, we perform a best fit to the Sachs electric 
and magnetic form factors, as they are the usual form factors extracted 
from experiment.  
The Sachs electric and magnetic form factors are determined  
from $F_1$ and $F_2$ as
\begin{eqnarray}
G_E(t) &\,=\,& F_1(t) \,-\, \tau \, F_2(t) \;, \\ 
G_M(t) &\,=\,& F_1(t) \,+\, F_2(t) \;, 
\end{eqnarray} 
where $\tau \equiv -t / 4 M_N^2$.
\newline
\indent
The Regge slope parameter $\alpha^{\prime}$ 
can in principle be directly fitted from the knowledge of the 
electromagnetic radii of proton and neutron. In particular, 
the electric mean squared radii of proton and neutron are 
given by 
\begin{eqnarray}
r^2_{E, p} &\,=\,& r^2_{1, p}
\;+\; {3 \over 2} \, {{\kappa_p} \over {M_N^2}} \; ,
\label{eq:rmsep} \\
r^2_{E, n} &\,=\,& r^2_{1, n}
\;+\; {3 \over 2} \, {{\kappa_n} \over {M_N^2}} \; ,
\label{eq:rmsen}
\end{eqnarray}
where the first term on the {\it rhs} is the Dirac radius squared $r^2_1$, 
whereas the second term is the Foldy term. The Dirac radii are 
calculated through the integrals of Eqs.~(\ref{eq:rms1p},\ref{eq:rms1n}) 
for the R1 model, and through  Eqs.~(\ref{eq:rms1p_2},\ref{eq:rms1n_2}) 
for the R2 model.  
\begin{figure}[h]
\includegraphics[width=10cm]{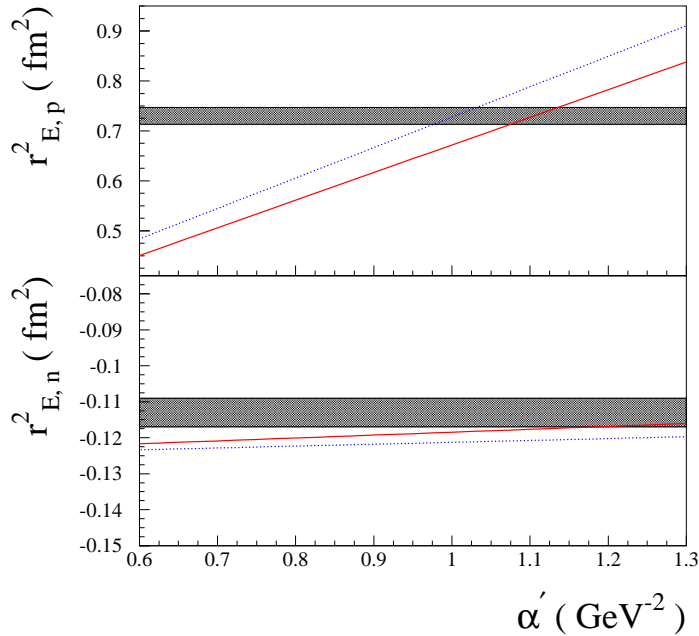}
\caption[]{\small Proton and neutron electric mean squared radii 
$r^2_{E, p}$ (upper panel) and $r^2_{E, n}$ (lower panel),   
Eqs.~(\ref{eq:rmsep},\ref{eq:rmsen}). 
Dotted curves: Regge ansatz 
according to Eqs.~(\ref{eq:rms1p},\ref{eq:rms1n}); 
solid curves : modified Regge ansatz  
according to Eqs.~(\ref{eq:rms1p_2},\ref{eq:rms1n_2}).  
Both calculations are shown as   
function of the Regge slope $\alpha^{'}$.  
For the quark distributions, 
the MRST02 NNLO parametrization \cite{Martin:2002dr} 
at scale $\mu^2$ = 1 GeV$^2$ was used in the calculations.
The shaded bands correspond to the experimental values. 
Note that for the neutron, the Foldy term (term proportional to
$\kappa_n$ in Eq.~(\ref{eq:rmsen}))
gives $r^2_{E, n}$ = - 0.126 fm$^2$.}
\label{fig:rmse}
\end{figure}
\newline
\indent
In Fig.~\ref{fig:rmse}, we show the proton and neutron rms radii as the 
functions of the Regge slope $\alpha^{\prime}$ for both R1 and R2 
models.  
One notes that the neutron rms radius is dominated by 
the Foldy term, 
which gives $r^2_{E, n}$ = - 0.126 fm$^2$. Therefore, a relatively
wide range of values $\alpha^{'}$ are compatible with the neutron
data. However for the proton, a rather narrow range of values around 
$\alpha^{'} = 1.0 - 1.1$~GeV$^{-2}$ are favored. Such value is close
to the expectation from the near universal Regge slopes for 
meson trajectories, therefore supporting 
our Regge type parametrizations.   
\begin{figure}[h]
\includegraphics[width=10cm]{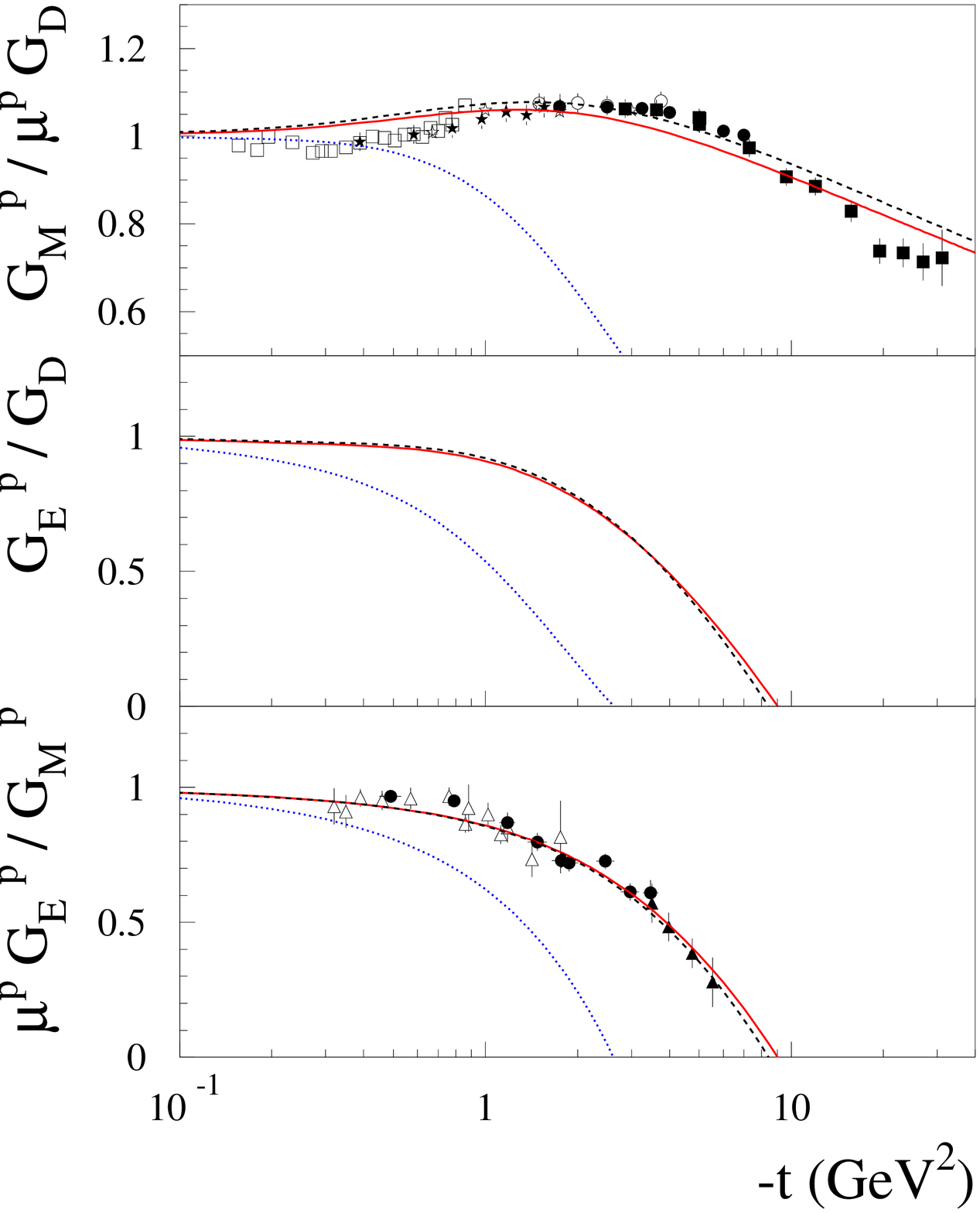}
\vspace{0cm}
\caption[]{\small Proton magnetic (upper panel) and electric (middle
  panel) form factors relative to the dipole form 
$G_D(t) = 1/(1 - t/0.71)^2$, as well as 
the ratio of both form factors (lower panel). 
The dotted curves correspond to the  Regge parametrization R1, 
with $\alpha^{'}= 1.105 \,$ GeV$^{-2}$.  
The solid and dashed curves correspond to two fits using the 
modified Regge  parametrization R2.  
The solid curves are for the 3 parameter fit : 
$\alpha^{'} = 1.105$\, GeV$^{-2}$, $\eta_u$ = 1.713 and $\eta_d$ = 0.566. 
The dashed curves are for the 2 parameter fit : 
$\alpha^{'} = 1.09$\, GeV$^{-2}$, $\eta_u = \eta_d$ = 1.34.   
Data for the proton magnetic form factor $G_M^p$ are from   
\cite{Janssens66} (open squares), \cite{Litt70} (open circles),
\cite{Berger71} (solid stars), \cite{Bartel:1973rf} (open stars), 
\cite{Andivahis:1994rq} (solid circles), \cite{Sill:1992qw} 
(solid squares), 
according to the recent re-analysis of Ref.~\cite{Brash:2001qq}. 
Data for the ratio $G_E^p / G_M^p$ are from \cite{Jones:1999rz} (solid
circles), \cite{Gayou:2001qt} (open triangles), 
and \cite{Gayou:2001qd} (solid triangles).}
\label{fig:gegmproton}
\end{figure}
\begin{figure}[h]
\includegraphics[width=10cm]{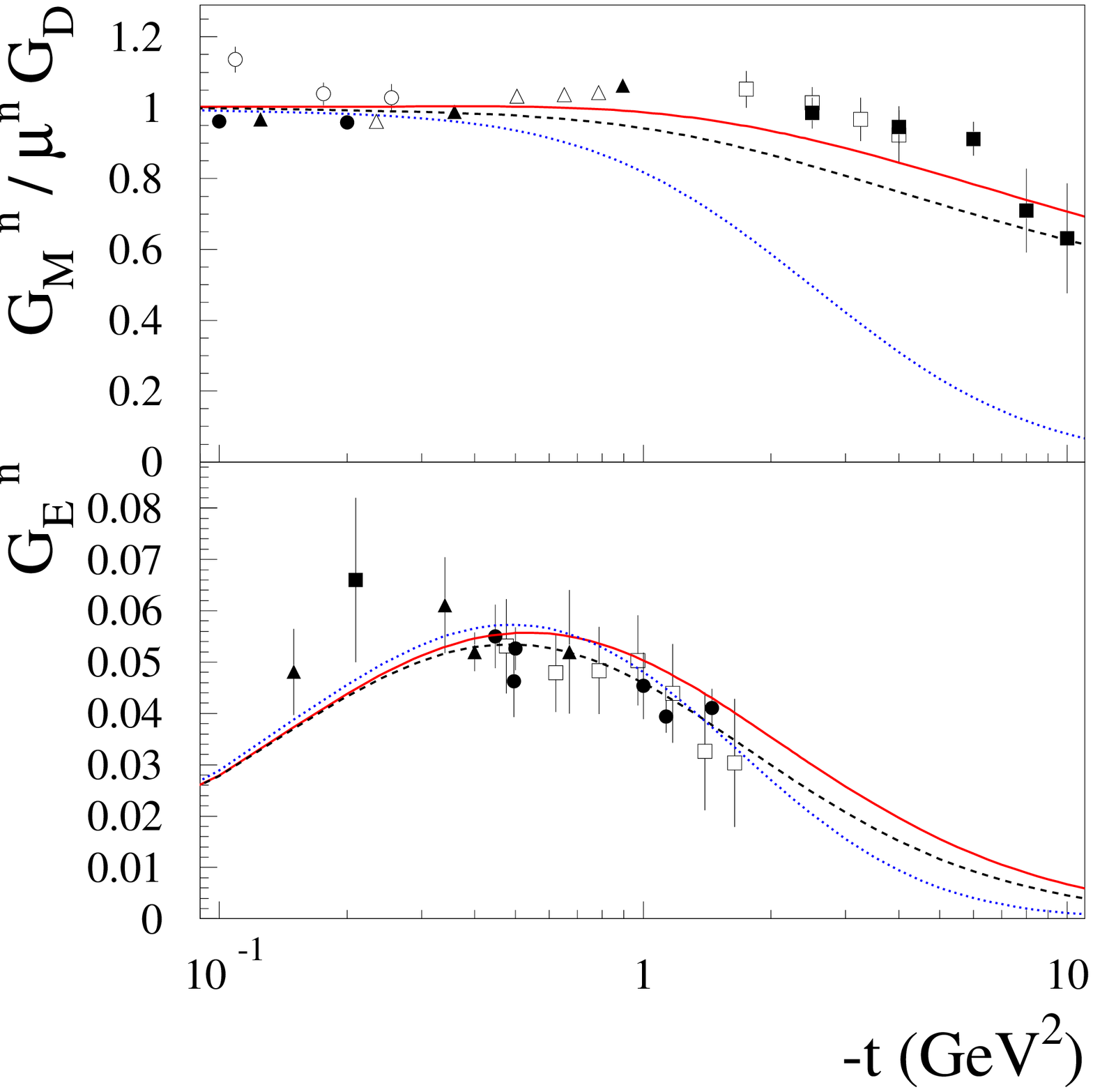}
\vspace{0cm}
\caption[]{\small Neutron magnetic form factor relative to the dipole form 
(upper panel), and neutron 
electric form factor (lower panel), with curve conventions 
as in Fig.~\ref{fig:gegmproton}. 
The data for the neutron magnetic form factor $G_M^n$ are from 
\cite{Markowitz:hx} (open circles), \cite{Xu:2000xw} (solid circles),   
\cite{Anklin:ae} (open triangels), \cite{Kubon:2001rj} (solid triangles), 
\cite{Lung:1992bu} (open squares), and \cite{Rock:1982gf} (solid squares).
The data for the neutron electric form factor $G_E^n$ are from different
double polarization experiments at MAMI (triangles 
\cite{Herberg:ud,Ostrick:xa,Becker:tw,Rohe:sh}), 
NIKHEF (solid square \cite{Passchier:1999cj}) 
and JLab (solid circles \cite{Zhu:2001md,Warren:2003ma,Madey:2003av}). 
The open squares are the $G_E^n$ extraction from the 
deuteron quadrupole form factor according to 
the analysis of Ref.~\cite{Schiavilla:2001qe}.}
\label{fig:gegmneutron}
\end{figure}
\newline
\indent
In Figs.~\ref{fig:gegmproton}, \ref{fig:gegmneutron}, we show the 
proton and neutron Sachs electric and magnetic form factors. 
One observes from Figs.~\ref{fig:gegmproton}, \ref{fig:gegmneutron} 
that the modified Regge model R2 gives a rather good description of all 
available form factor data for both proton and neutron in the whole $t$ range 
using the parameter for the Regge trajectory 
$\alpha^{'}$ = 1.105 \,GeV$^{-2}$, 
and   the following values for the coefficients 
governing the $x \to 1$ behavior of the $E$-type GPDs:  
$\eta_u$ = 1.713 and $\eta_d$ = 0.566.  
The 2-parameter version of the R2 model gives a description of similar quality 
if we take $\alpha^{'}$ = 1.09 \,GeV$^{-2}$ and $\eta_u= \eta_d = 1.34$. 
\newline
\indent
In Figs.~\ref{fig:gegmproton} and \ref{fig:gegmneutron}, we also show 
the results of the initial  Regge model R1, 
with the above value $\alpha^{'}$ = 1.105 \,GeV$^{-2}$ of 
$\alpha^{\prime}$. 
One sees from Figs.~\ref{fig:gegmproton}, \ref{fig:gegmneutron} that 
the Regge model R1 is able to reproduce the main trends of both proton and 
neutron electromagnetic form factor data for $-t \leq 0.5$~GeV$^2$. 
For higher values of $-t$, however, it falls short of the data,
since as we discussed,  it predicts  faster power fall-off 
than that corresponding to the DY relation. 
The modified Regge model R2  reproduces the DY powers for the
form factors at large $-t$, and   is able to accurately 
describe existing data.
The two additional parameters $\eta_u$ and $\eta_d$ in the R2 model, 
in particular, allow 
to describe the decreasing ratio of $G_E^p / G_M^p$ with increasing momentum 
transfer, as follows from the recent JLab polarization 
experiments~\cite{Jones:1999rz,Gayou:2001qt,Gayou:2001qd}.  
Our parametrization leads to a zero for $G_E^p$ at a 
momentum transfer of $-t \simeq 8$~GeV$^2$, which will be within the range 
covered by an upcoming JLab experiment~\cite{E-01-109}. 
\begin{figure}[ht]
\includegraphics[width=10cm]{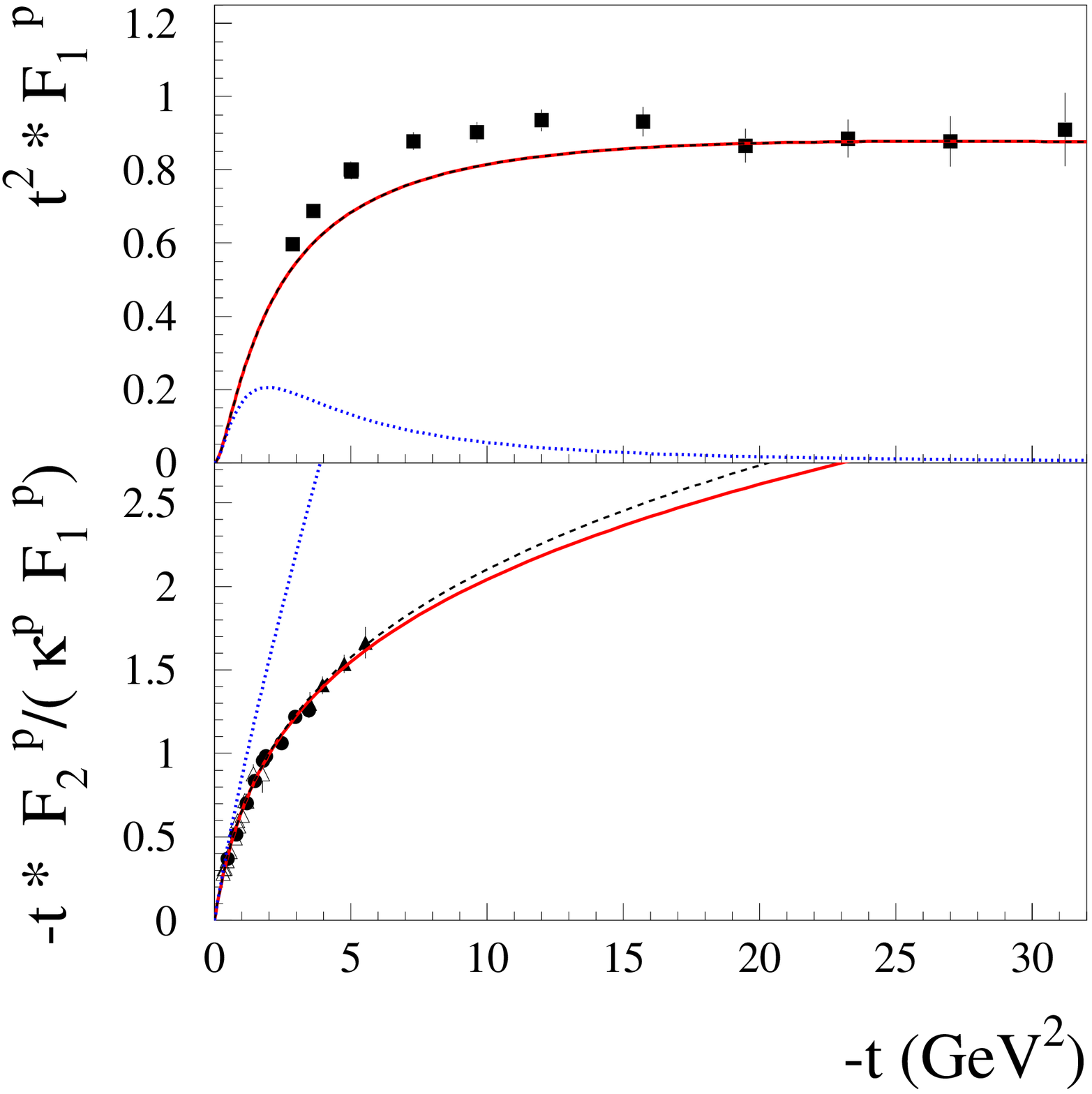}
\vspace{0cm}
\caption[]{\small Proton Dirac form factor (upper panel) 
multiplied by $t^2$ 
and ratio of Pauli to Dirac form factor multiplied by $-t$ (lower panel), 
with curve conventions as in Fig.~\ref{fig:gegmproton}. 
The data for $F_1^p$ are from \cite{Sill:1992qw} (solid squares).  
Data for the ratio $F_2^p / F_1^p$ are from \cite{Jones:1999rz} (solid
circles), \cite{Gayou:2001qt} (open triangles), 
and \cite{Gayou:2001qd} (solid triangles).}
\label{fig:f1f2proton}
\end{figure}
\newline
\indent
To study the large $-t$ behavior of our GPD parametrizations, it is 
instructive to plot the Dirac and Pauli form factors. In this way, 
one separates the large $-t$ behavior of both the GPDs $H$ and $E$.  
In Fig.~\ref{fig:f1f2proton}, 
we show this large $-t$ behavior for $F_1^p$, and for 
the ratio of $F_2^p / F_1^p$. One observes from Fig.~\ref{fig:f1f2proton} 
that for $F_1^p$, the Regge parametrization R2 settles to an approximate 
$\sim 1/t^2$ power behavior around $-t \simeq 10$~GeV$^2$.

The ratio $F_2^p / F_1^p$ was also discussed within 
the context of  perturbative QCD (pQCD),
where the asymptotic large-$t$ behavior 
of the nucleon form factors is dominated by
diagrams with two hard gluon exchanges~\cite{Brodsky:1973kr,Brodsky:1974vy}.
In any model with dimensionless quark-gluon 
coupling constant, 
 these  diagrams give   $F_1^p \sim 1/t^2$ ~\cite{Brodsky:1973kr}.
 Furthermore, for vector gluons,  the  quark helicity conservation  
 at the gluon vertex 
and dimensional counting  suggest the  extra $m^2/t$ 
suppression for the 
  $F_2^p$ form factor ~\cite{Brodsky:1973kr,Lepage:1979za},
 with $m$ being the quark mass or a nonperturbative 
 parameter coming from the baryon wave function corresponding 
 to extra unit of orbital angular momentum~\cite{Belitsky:2002kj}.    
Thus, one should expect that 
$F_2 / F_1 \sim 1/t$ in pQCD.  
Direct calculation~\cite{Belitsky:2002kj}, however,  
shows that  the integrals over the quark momentum 
fractions $x_i,y_j$ in the pQCD formula 
contain terms like $\varphi (x_i, \ldots)
\varphi (y_j, \ldots)/x_i^2y_j^2$
that diverge even if the nucleon distribution 
amplitudes $\varphi (x_i, \ldots), \varphi (y_j, \ldots)$ linearly 
vanish at small $x_i,y_j$. 
Strictly speaking, this means that pQCD factorization 
is not applicable to calculating $F_2^p(t)$ even in the asymptotic 
$-t \to \infty$ limit, the fact well known since the pioneering 
papers~\cite{Lepage:1979za,Lepage:1980fj}.
The authors of Ref.~\cite{Belitsky:2002kj} substituted 
the logarithmic divergences by $\log(-t/\Lambda_{QCD}^2)$
factors, and    obtained 
$F_2^{\rm pQCD} / F_1^{\rm pQCD} \sim \log^2(-t/\Lambda_{QCD}^2) / (-t)$.
This result was found to be in surprisingly  good agreement with the JLab data.
In this connection, we want to
emphasize that 
our results for $F_2(t)$  and $F_1(t)$ correspond to the Feynman mechanism, 
i.e., to  overlap of soft wave functions.
The pQCD terms correspond to two iterations
of the soft wave functions with hard gluon
exchange kernels.
As is well known, there is ${\cal O} (\alpha_s/\pi)$ 
suppression for each extra loop of a Feynman diagram in QCD.
Thus, from our point of view,
pQCD terms are ${\cal O}((\alpha_s/\pi)^2)$ or,  at most,  a few per cent 
corrections to the Feynman mechanism
contributions to  $F_1$ and $F_2$. For this reason, we neglect them in our analysis.
In our  parametrization R2, the good description found for the 
ratio $F_2^p / F_1^p$ can be directly assigned to the extra suppressing  
factor of  $(1-x)^{\eta}$ 
contained in the GPD $E(x,t)$. 
The question, how this suppression is related to
the quark orbital angular momentum, deserves further 
investigation. It is interesting to note that the 
 extra $(1-x)$ factor for  ${\cal E}^u (x)$ function compared 
 to ${\cal H}^u (x)$ appears in the starting term of   the QCD sum rule 
 calculation of these functions \cite{Radyushkin:2004mt}.
  Also, the dominant $x \to 1$ perturbative QCD term for the GPD  $E$ 
 (given by $\alpha_s^4$ diagrams) involves two additional powers in
$(1 - x)$ compared with the pQCD expression for the leading $x \to 1$ 
term in the GPD $H$ ~\cite{Yuan:2003fs}. 
 \newline
\indent
Since  the GPD $E$ enters the sum rule 
for the total angular momentum $J^q$ carried by a quark of flavor $q$ 
in the proton as~\cite{Ji:1996ek}~:
\begin{equation}
2 J^q \,=\, \int_{-1}^{1} dx \, x \, 
\left\{ H^q(x, 0, 0) + E^q(x, 0, 0) \right\} ,
\label{eq:jisr}
\end{equation} 
our parametrization R2, 
in which the $x \to 1$ limit of $E$ 
is determined from the $F_2^p / F_1^p$ form 
factor ratio, allows to evaluate the above sum rule. 
The first term in the sum rule of Eq.~(\ref{eq:jisr})
is already known from the forward parton distributions and is equal to the 
total fraction of the proton momentum carried by a quark of flavor 
$q$ $(q = u, d, s)$~:
\begin{eqnarray}
M_2^q \,&\equiv&\, \int_{-1}^1dx \, x \, H^q(x, 0, 0), \nonumber \\
&=&\, 
\int_0^1dx \, x \, \left[ \, q_v(x) + 2 \, \bar q(x) \, \right] \, ,
\label{eq:m2}
\end{eqnarray}
with $\bar q(x)$ the anti-quark distribution. For the 'non-trivial' 
contribution  to the sum rule, arising from the second moment of the GPD $E$, 
we use our modified Regge parametrization R2 
of Eq.~(\ref{eq:e2}) for ${\cal E}^q(x)$, which,
neglecting the antiquark contribution,  yields for 
Eq.~(\ref{eq:jisr})~:
\begin{eqnarray}
2 \, J^u \,&=&\, M_2^u \,+\, \frac{\kappa^u}{N_u} \;
 \int_0^1dx \, x \, (1 - x)^{\eta_u} \, u_v(x)  \, , 
\label{eq:ju} \\
2 \, J^d \,&=&\, M_2^d \,+\, \frac{\kappa^d}{N_d} \;
 \int_0^1dx \, x \, (1 - x)^{\eta_d} \, d_v(x)   \, , 
\label{eq:jd} \\
2 \, J^s \,&=&\, M_2^s \, .
\label{eq:js}
\end{eqnarray}
\begin{table}[h]
{\centering \begin{tabular}{|c|c|c|c|}
\hline
&&& \\ 
&$M_2^{q}$ (MRST2002) & $2 \, J^q$ (R2 model) & $2 \, J^q$ 
(lattice \cite{Gockeler:2003jf}) \\
&&& \\
\hline 
\hline
$u$ & 0.37 & 0.58 & 0.74 $\pm$ 0.12 \\
\hline 
$d$ & 0.20 & -0.06 & -0.08 $\pm$ 0.08 \\
\hline 
$s$ & 0.04 & 0.04 &  \\
\hline 
$u + d + s$ & 0.61 & 0.56 & 0.66 $\pm$ 0.14 \\
\hline 
\hline
\end{tabular}\par}
\caption{\small Estimate of $2 \, J^q$ (second column) 
for the different quark flavors at the scale $\mu^2$ = 2 GeV$^2$ 
according to Eqs.~(\ref{eq:ju}-\ref{eq:js}),  
using the R2 parametrization (with 3 parameters) for the GPD $E$. 
For the forward parton distributions, 
the MRST2002 NNLO parametrization~\cite{Martin:2002dr} is used, 
yielding the total quark momentum contributions $M_2^q$ (first column). 
For comparison, the third column shows the quenched lattice QCD results 
of~\cite{Gockeler:2003jf}, extrapolated to the physical pion mass, 
for $2 \, J^u$ and $2 \, J^d$. 
\label{table_spin}}
\end{table}

In Table~\ref{table_spin}, we show the 
values of the quark momentum sum rule $M_2^q$ at the
scale $\mu^2$ = 2 GeV$^2$, using the MRST2002 
parametrization~\cite{Martin:2002dr} for the forward parton distributions.
We also show the estimate for $J^u$, $J^d$, and $J^s$ of 
Eqs.~(\ref{eq:ju}-\ref{eq:js})  at the same scale. 
As was already observed in Ref.~\cite{Goeke:2001tz}, 
based on a Regge model of the type R1, 
our estimates lead to a large fraction (63 \%) of the total angular
momentum of the proton carried by the $u$-quarks and a relatively
small contribution carried by the $d$-quarks. 
As the $d$-quark intrinsic spin contribution is known to be relatively large 
and negative ( $\Delta d_v \simeq -0.25$ ), the small total angular momentum 
contribution $J^d$ of the $d$-quarks which follows from our parametrization 
implies an interesting cancellation between the intrinsic spin contribution 
and the orbital contribution $L^d$ ( with $2 J^q = \Delta q + 2 L^q$ ), 
which should therefore be of size $2 L^d \simeq 0.2$. For the $u$-quark on the 
other hand,  the parametrization R2 yields only a small value 
for $2 L^u$, as our estimate for $2 J^u$ is quite close to 
the intrinsic spin contribution $\Delta u_v \simeq 0.6$. 
Such a picture is also supported by a recent quenched lattice QCD 
calculation~\cite{Gockeler:2003jf} (see also \cite{Mathur:1999uf} for an 
earlier calculation) 
for the valence quark contributions to $2 J^u$ and $2 J^d$. 
One indeed sees from Table~\ref{table_spin} (third column) that the 
quenched lattice QCD calculation yields quite similar values for 
$2 J^u$ and $2 J^d$ as our parametrization R2. 
It remains to be seen however how large is the sea quark 
contribution to the GPD $E$ which can enter the spin sum rule of 
Eq.~(\ref{eq:jisr}). This sea quark contribution is 
only approximately included 
(i.e. fermion loop contributions are neglected) 
in the quenched lattice QCD calculations of Ref.~\cite{Gockeler:2003jf}.  
An exploratory investigation using  
unquenched QCD configurations has been performed in Ref.~\cite{Hagler:2003jd}.
The sea quark contribution is also not constrained by the form factor 
sum rules considered in this paper, which only constrain the valence quark 
distributions. Ongoing measurements 
of hard exclusive processes, such as deeply virtual Compton scattering, 
provide a means to address this question in the near future.

Besides the electromagnetic form factors for proton and neutron, 
the Regge parametrizations discussed in this work can also be used to estimate 
$N \to \Delta$ transition form factors, provided one can relate the 
$N \to \Delta$ transition GPDs to the $N \to N$ ones. 
First experiments which are sensitive to the $N \to \Delta$ GPDs 
have recently been reported~\cite{Guidal:2003ji}.
For the magnetic $N \to \Delta$ transition form factor $G_M^{*}(t)$, 
it was shown in Ref.~\cite{Frankfurt:1999xe} that, 
in the large $N_c$ limit, the 
relevant  $N \to \Delta$ GPD can be expressed in terms of the
isovector GPD $E^u - E^d$, yielding the sum rule   
\begin{eqnarray}
G_M^{*}(t) \,=\, {{G_M^{*}(0)} \over {\kappa_V}} \; \int _{-1}^{+1}dx\; 
\biggl \{ E^{u}(x,\xi ,t) \,-\, E^{d}(x,\xi ,t) \biggr \} 
\; =\;  {{G_M^{*}(0)} \over {\kappa_V}} \; 
\biggl \{ F_2^p(t) - F_2^n(t) \biggr \} \, ,  
\label{eq:gmsumrule} 
\end{eqnarray}
where $\kappa_V = \kappa_p - \kappa_n = 3.70$. 
Within the large $N_c$ approach used in  Ref.~\cite{Frankfurt:1999xe},
the value $G_M^{*}(0)$ is  given by  
$G_M^{*}(0) = \kappa_V / \sqrt{3}$~\cite{Goeke:2001tz}, 
which  is about  30\% smaller than the
experimental number. In our calculations, we will therefore use the
phenomenological value $G_M^{*}(0) \approx 3.02$ \cite{Tiator:2000iy}. 
\begin{figure}[ht]
\includegraphics[width=10cm]{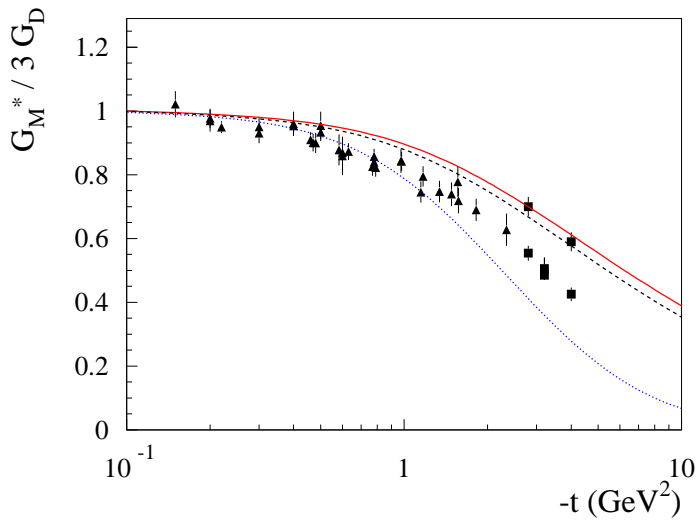}
\vspace{0cm}
\caption[]{\small The $N \to \Delta$ magnetic 
transition form factor, relative to the 
dipole form (multiplied by a factor 3). 
Curve conventions as in Fig.~\ref{fig:gegmproton}.
The data for $G_M^{*}$ are from the compilation of~\cite{Tiator:2000iy}. 
For the JLab data points at 2.8 and 4 GeV$^2$, 
both the analyses of \cite{Frolov:1998pw} (upper points) and
\cite{Tiator:2000iy} (lower points) are shown.}
\label{fig:gmdel}
\end{figure}
\newline
\indent
We show our results for $G_M^*$ using Eq.~(\ref{eq:gmsumrule}) in 
Fig.~\ref{fig:gmdel}. It is seen that both the Regge and modified Regge 
parametrizations yield a magnetic $N \to \Delta$ form factor which decreases 
faster than a dipole, in qualitative good agreement with the data. 

The sum rule (\ref{eq:gmsumrule}) was 
used earlier by P. Stoler
\cite{Stoler:2002im},  who proposed a  model 
\cite{Stoler:2001xa,Stoler:2003mx}  in which
the Gaussian ansatz  for GPDs is  modified at large $-t$ by  terms having 
a power-law behavior.

\section{GPDs in impact parameter space and positivity constraints}
\label{sec:impact}

The models for GPDs should satisfy many constraints.
In fact, such constraints as 
the reduction of GPDs to usual parton densities in the forward limit
and to form factors in the local  limit, 
are the key points for the models  constructed in this paper.
There are more complicated constraints  imposed, e.g.,  by the polynomiality
condition  which is extremely important 
for nonzero skewness. Since  the nonforward parton densities 
correspond to   $\xi = 0$, they   are not affected 
by these constraints. However, they are affected by    the positivity conditions 
which should be taken into account 
both for nonzero and zero skewness parameter. 
In particular, there exists a relation between the
 $E$-type and $H$-type   GPDs \cite{Burkardt:2003ck}.
  Since we are constructing $E$-GPDs from $H$-GPDs by
  a simple modification of the $x$-behavior of $H$ by a power
  of $(1-x)$, we should check that such a modification 
  is consistent  with the positivity constraint of Ref. \cite{Burkardt:2003ck}. 

The most convenient formulation  of the 
positivity constraint relating the $E$ and $H$ GPDs 
is  in the impact parameter space.  
For $\xi = 0$, the  impact parameter versions of GPDs are obtained 
through a Fourier integral 
in transverse momentum $q_\perp$~:
\begin{eqnarray}
H^q(x, {\bf b_\perp}) \,& \equiv &\, 
\int \frac{d^2 {\bf q_\perp}}{(2 \pi)^2} \, 
e^{i {\bf b_\perp \cdot q_\perp}} \;
{\cal H}^q (x, - {\bf q_\perp^2}) , \\ 
E^q(x, {\bf b_\perp}) \,& \equiv &\, 
\int \frac{d^2 {\bf q_\perp}}{(2 \pi)^2} \, 
e^{i {\bf b_\perp \cdot q_\perp}} \;
{\cal E}^q (x, - {\bf q_\perp^2}) ,  
\end{eqnarray}
These functions   have the physical meaning of measuring the 
probability to find a quark which carries longitudinal 
momentum fraction $x$ at a transverse position ${\bf b_\perp}$ in a nucleon, 
see Refs.~\cite{Burkardt:2000za,Burkardt:2002hr}.

It has been shown \cite{Burkardt:2003ck} that the GPDs $H$ and $E$ in the impact 
parameter space satisfy the positivity bound~:
\begin{eqnarray}
\frac{1}{2 M_N} \, \left| {\bf \nabla_{b_\perp}} \, E^q(x, {\bf b_\perp}) 
\right| 
\,& \leq &\, H^q(x, {\bf b_\perp}). 
\label{eq:positiv}
\end{eqnarray}

\begin{figure}[h]
\includegraphics[width=12cm]{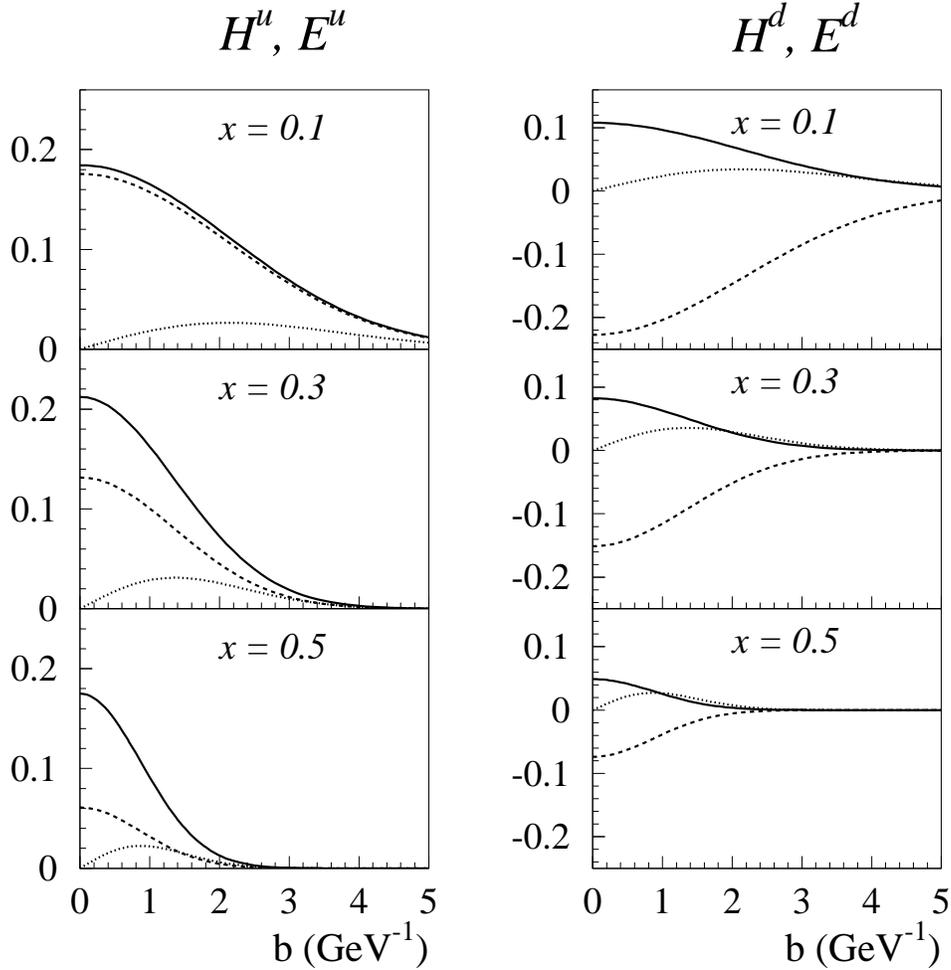}
\vspace{0cm}
\caption[]{\small GPDs in impact parameter space for the modified 
Regge parametrization R2,  
with parameters 
$\alpha^{'} = 1.105$\, GeV$^{-2}$, 
$\eta_u$ = 1.713 and $\eta_d$ = 0.566.
Left panels for $u$-quark; right panels for $d$-quark. 
The solid (dashed) curves give  GPDs $H^q$ ($E^q$), respectively. 
The dotted curves correspond to the function 
$\left| {\bf \nabla_{b_\perp}} \, E^q \right| / (2 M_N)$ entering 
the positivity bound of Eq.~(\ref{eq:positiv}).  
}
\label{fig:gpdimpact}
\end{figure}

\begin{figure}[h]
\includegraphics[width=10cm]{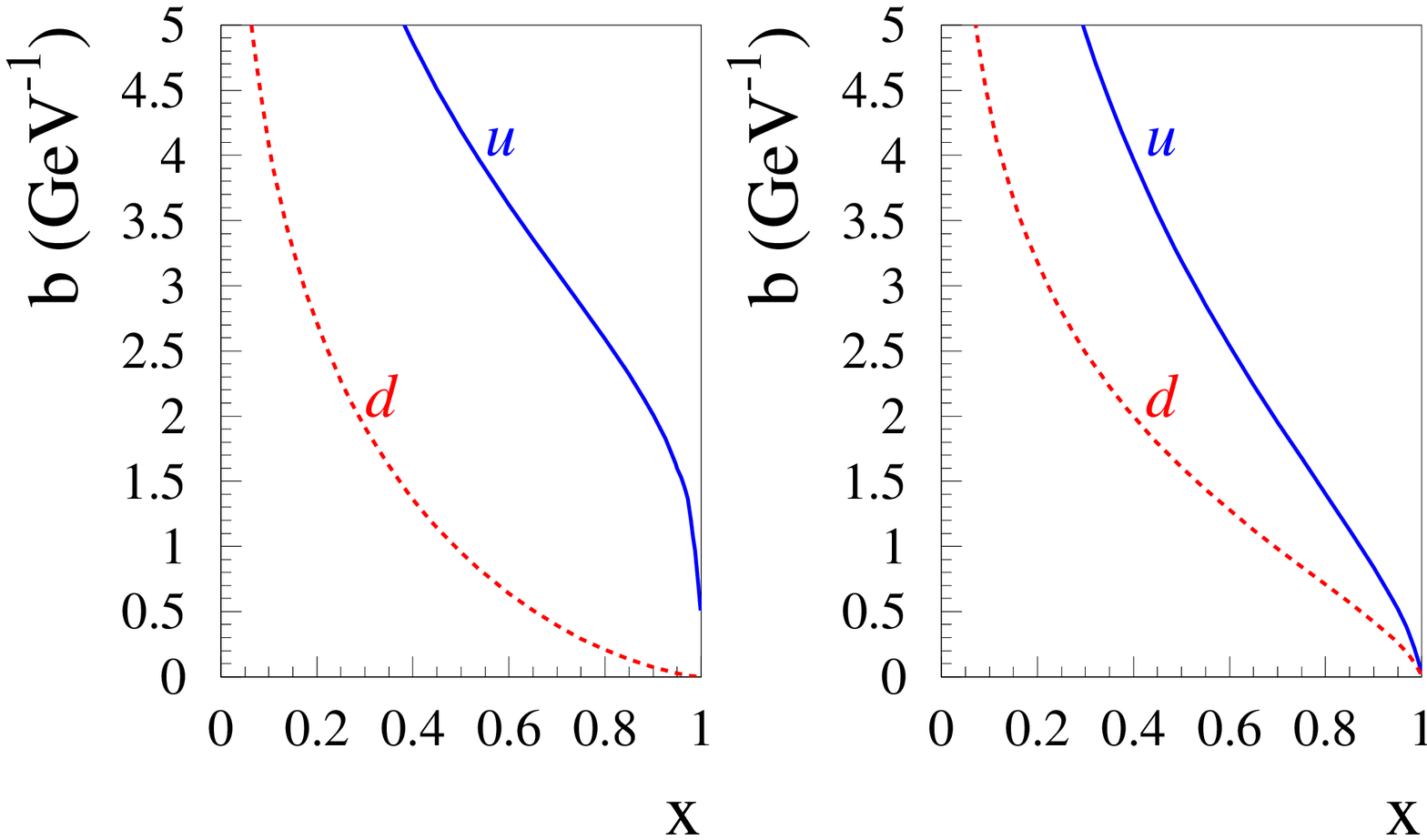}
\vspace{0cm}
\caption[]{\small Upper bound on ${\bf b_\perp}$ ( Eq.~(\ref{eq:upperb}) ) 
required by the positivity bound of Eq.~(\ref{eq:positiv}) 
for two fits using the Regge parametrization R2. 
Left panel is for the 3 parameter fit : $\alpha^{'} = 1.105$\, GeV$^{-2}$, 
$\eta_u$ = 1.713 and $\eta_d$ = 0.566.
Right panel is for the 2 parameter fit : $\alpha^{'} = 1.09$\, GeV$^{-2}$, 
$\eta_u = \eta_d$ = 1.34.   
}
\label{fig:upperb}
\end{figure}

Translating  the GPD parametrization R2 of Eqs.~(\ref{eq:hr2},\ref{eq:er2}), 
into the  impact parameter space, we obtain~:
\begin{eqnarray}
H^q (x, {\bf b_\perp}) \,&=&\, q_v(x) \, 
\frac{e^{- {\bf b_\perp}^2 \,/\, [ - 4 \, \alpha^{\, '} \, 
(1-x) \ln x ]}}{4 \pi \, \left[-\alpha^{\, '} (1-x) \ln x  \right]}  , 
\\
E^q (x, {\bf b_\perp}) \,&=&\, \frac{\kappa_q}{N_q} \, (1 - x)^{\eta_q} 
\, q_v(x) \, 
\frac{e^{- {\bf b_\perp}^2 \,/\, [ - 4 \, \alpha^{\, '} \,
 (1-x) \ln x ]}}{4 \pi \, \left[-\alpha^{\, '} (1-x) \ln x  \right]}  , 
\label{eq:her2b}
\end{eqnarray}
from which it follows that
\begin{eqnarray}
\left| {\bf \nabla_{b_\perp}} \, E^q(x, {\bf b_\perp}) 
\right| \,=\, \frac{\kappa_q}{N_q} \, (1 - x)^{\eta_q} 
\, q_v(x) \, \frac{\left| \bf{b_\perp} \right|}{2} \,
\frac{e^{- {\bf b_\perp}^2 \,/\, [ - 4 \, \alpha^{\, '} \, 
(1-x) \ln x ]}}{4 \pi \, \left[-\alpha^{\, '} (1-x) \ln x  \right]^2}  , 
\label{eq:der2b}
\end{eqnarray}
Within the R2 parametrization, the positivity bound of Eq.~(\ref{eq:positiv}) 
implies an upper bound on the value of $\left| \bf {b_\perp} \right|$~:
\begin{eqnarray}
\left| \bf {b_\perp} \right| \,\leq\, \frac{N_q}{\left| \kappa_q \right|} \, 
\frac{M_N}{(1 - x)^{\eta_q}} \, 
4\, \left[-\alpha^{\, '} (1-x) \ln x  \right]  . 
\label{eq:upperb}
\end{eqnarray}
\newline
\indent
In Fig.~\ref{fig:gpdimpact}, 
we show the GPDs in the impact parameter space for the modified 
Regge parametrization R2 discussed above. 
The parameters are taken from the best fit to the form factors 
as discussed in the previous section.  
We see from Fig.~\ref{fig:gpdimpact} that for the $u$-quark GPDs, the 
positivity bound of Eq.~(\ref{eq:positiv}) is satisfied over most of the 
$x$-region, considering that the GPDs are vanishingly small for 
values of ${\bf b_\perp}$ larger than the nucleon size (corresponding with 
about 4.4~GeV$^{-1}$).  
For the $d$-quark GPDs on the other hand, there is a violation 
in the present parametrization, which becomes more pronounced at 
larger values of $x$ and ${\bf b_\perp}$, as is shown 
in Fig.~\ref{fig:upperb} (left panel). 
We therefore tried to extend the range of validity of the R2 parametrization 
by finding a fit with a higher value of $\eta_d$. This can be obtained 
by imposing the constraint $\eta_u = \eta_d$. We have shown before that the 
resulting two-parameter fit 
($\alpha^{'} = 1.09$\, GeV$^{-2}$, $\eta_u = \eta_d$ = 1.34) 
gives a nearly as satisfactory description of the form factors. 
It is seen from the right panel in Fig.~\ref{fig:upperb} 
that this 2-parameter fit extends the region in $x$ and ${\bf b_\perp}$ for 
the d-quark where our parametrization satisfies the positivity condition. 

It is clear, that with a somewhat  more complicated 
model, we can easily satisfy the positivity constraint.
However, given that the violation is rather small,
we prefer not to introduce extra parameters 
and to  keep the parametrization as simple as possible.

Furthermore, it is clearly seen from these images that  for  large values of 
$x$,   
our  quark distributions  are concentrated at small values of ${\bf b_\perp}$, 
reflecting the distribution of valence quarks in the core of the 
nucleon. On the other hand, at small values of $x$, 
the distribution in transverse position extends much further out.  
This expected correlation  assures that our model
correctly reproduces the gross features of the nucleon structure
as expressed in terms of the quark distributions.

\section{Conclusions}
\label{sec:conclusions}

Summarizing, we discussed in this work several parametrizations for the 
$t$-dependence of the nucleon GPDs in view of the recent accurate data 
for the nucleon electromagnetic form factors in the spacelike region. 
\newline
\indent
Starting from the low $-t$ region, we discussed a Regge model in 
which the $x$ and $t$ dependence of the GPDs are coupled in the 
form $x^{- \alpha^\prime \, t}$. This model has only one  parameter which 
physically corresponds to  the slope $\alpha^\prime$  
of the Regge trajectory  in the 
vector EM current channel. 
 This  parameter is  linearly related to the rms 
radii of $F_1$ and $F_2$ form factors, and it  was  found 
that both radii are well deccribed by the same  universal Regge slope. 
\newline
\indent
Such a Regge model leads however to  faster power fall-off of  
 form factors in the large $-t$ region than that expected
 from the Drell-Yan relation. 
To conform with this relation and the observed power behavior at large $-t$,
we used a modified Regge parametrization  that 
gives slower decrease with $-t$. 
The modified Regge parametrization 
displays approximately a $1/t^2$ behavior for $F_1^p(t)$ data in the region 
$-t \geq 10$~GeV$^2$.  To describe  $F_2^p(t)$, we need to introduce, 
 in addition to   $\alpha^\prime$, two parameters that govern the $x \to 1$ 
behavior of the GPD $E$.  They were adjusted  to  
give an accurate description of 
the recent polarization data for 
the ratio $F_2^p / F_1^p$. 
Since  this behavior in our model is correlated with the 
$x \to 1$ behavior of the GPD $E$, it also allows us to evaluate the 
sum rule for the total angular momentum carried by the quarks, 
which involves the second moment of the GPD $E$. 
\newline
\indent
For the quark contributions to the nucleon spin, we find an intriguing 
flavor dependence, in which the valence $u$-quark contributes about 
two-thirds of the proton's spin (at a low renormalization point), which is 
nearly entirely arising from the $u$-quarks intrinsic spin contribution. 
For the valence $d$-quark on the other hand, our parametrization implies 
a near cancellation between its negative intrinsic spin contribution 
and its orbital angular momentum contribution. Recent quenched lattice 
QCD calculations support this observation. 
\newline
\indent
It remains to be seen by how much the sea quarks affect this picture. 
Ongoing measurements of hard exclusive processes, such as deeply 
virtual Compton scattering, are a means to address this question. 
As the GPDs mostly enter in hard exclusive observables through 
convolution integrals, our parametrization, which builds in the 
constraint coming from the first moment through the nucleon 
electromagnetic form factors, can be used as a first step to 
unravel the information on GPDs from the observables. 
The present work also suggests several interesting directions for future 
research. One of them is the extension of this study to 
quantify the link between the nucleon strangeness 
form factors and the $s$-quark distributions. Furthermore, 
the study of the chiral corrections (pion mass dependence) to the 
GPDs will allow to match onto the corresponding known chiral behavior of the 
elastic form factors at small momentum transfer.

\section*{Acknowledgements}

This work  is supported by the US 
 Department of Energy  contract
DE-AC05-84ER40150 under which the Southeastern
Universities Research Association (SURA)
operates the Thomas Jefferson Accelerator Facility;
by the  US Department of Energy grant DE-FG02-04ER41302 (M.V), 
by the French Centre National de la Recherche Scientifique (M.G.), 
and by the Alexander von Humboldt Foundation (M.P. and A.R.). 
The authors also like to thank the Institute for Nuclear Theory 
at the University of Washington, where part of this work was performed, 
for its hospitality.  One of us (A.R.) thanks S.J. Brodsky
for  correspondence about pQCD, Bloom-Gilman and Drell-Yan relations.

\end{document}